# Inducing magnetism onto the surface of a topological crystalline insulator

B. A. Assaf[1], F. Katmis[2,3], P. Wei[2], C.Z. Chang[2], B. Satpati[4], J.S. Moodera[2,3] and D. Heiman[1]

[1] Department of Physics, Northeastern University, Boston MA, 02115, USA
[2] Francis Bitter Magnet Lab, Massachusetts Institute of Technology, Cambridge MA, 02139, USA
[3] Department of Physics, Massachusetts Institute of Technology, Cambridge MA, 02139, USA
[4] Saha Institute of Nuclear Physics, 1/AF, Bidhannagar, Kolkata, 700064, India

**Abstract**

Inducing magnetism onto a topological crystalline insulator has been predicted to result in several novel quantum electromagnetic effects. This is a consequence of the highly strain-sensitive band topology of such symmetry-protected systems. We thus show that placing the TCI surface of SnTe in proximity to EuS – a ferromagnetic insulator – induces magnetism at the interface between SnTe and EuS and thus breaks time-reversal-symmetry in the TCI. Magnetotransport experiments on SnTe-EuS-SnTe trilayer devices reveal a hysteretic lowering of the resistance at the TCI surface that coincides with an increase in the density of magnetic domain walls. This additional conduction could be a signature of topologically-protected surface states within domain walls. Additionally, a hysteretic anomalous Hall effect reveals that the usual in-plane magnetic moment of the EuS layer is canted towards a perpendicular direction at the interface. These results are evidence of induced magnetism at the SnTe-EuS interfaces resulting in broken time-reversal symmetry in the TCI.

## I. INTRODUCTION

The behavior of Dirac fermions in topological surface states in proximity to magnetic films is poised to launch a new area of research.[1-11] This attention stems from several predicted quantized magnetoelectric phenomena[12-16] when a topological insulator (TI) is brought in proximity to a ferromagnet.[1,2,17] It has recently been established experimentally that an insulating ferromagnet is capable of inducing ferromagnetism by proximity onto the surface states of a TI, causing time-reversal-symmetry to be broken.[3,4,5,7] Furthermore, inducing magnetism onto topological surface states may lead to the quantized anomalous Hall effect (QAHE)[2,18] by virtue of broken time-reversal symmetry at the surface.[19-23] Such a proximity-induced QAHE is possible without requiring magnetic doping, such as in the case of Cr-doped $(Bi,Sb)_2Te_3$[24] and opens up the possibility of realizing a high-mobility QAHE. Of particular importance is the QAH state of a topological crystalline insulator (TCI), in which the Chern number could be tuned, up to values as high as four times that in standard single-Dirac-cone TI.[25] This is a consequence of the unique four-fold Dirac state degeneracy of TCI.[26-30] Inducing magnetism onto a TCI by proximity is thus a first important step towards this goal. The proximity method is advantageous as it does not rely on doping the bulk with magnetic transition metal ions[23,24,31], which would be expected to reduce the carrier mobility.

In this work, we investigated the magnetism induced onto the surfaces of the TCI SnTe by proximity to EuS, a ferromagnetic insulator (FMI), in MBE grown TCI/FMI/TCI trilayer structures. We observe a proximity-induced hysteretic negative magnetoresistance (MR) and an anomalous Hall effect – both evidence of magnetism induced at the interface between the non-magnetic TCI and the FMI. By examining the behavior of the resistance of the trilayer as a function of the magnetic domain density, it was found that the resistance minima at the coercive field of EuS, as well as the zero-field resistance of the TCI, are correlated with the magnetic domain texture of the FMI. These results agree with previous theoretical and experimental findings suggesting that the hysteretic MR arises as a result of domain-wall-trapped one-dimensional (1D) conduction channels at the TCI-FMI interface. [1, 3, 31] Moreover, the AHE and magnetization measurements indicate that the magnetic moment at the interfaces is canted, leading to an out-of-plane magnetic component, a necessary condition for proximity-induced time-reversal-symmetry breaking. [1,2,3,6,26]

## II. RESULTS AND DISCUSSION

Figure 1(a) illustrates the SnTe (14nm) / EuS (3nm) / SnTe (10nm) trilayer structure grown on Si (100). [32,33,34] SnTe [35] and EuS were both grown by e-beam evaporation from composite sources (see supplement S1 for details). [36-41] The structure of the trilayer was studied by X-ray diffraction and cross-sectional transmission electron microscopy (XTEM). Figure 1(b) shows an electron diffraction pattern of the trilayer. Diffraction spots corresponding to the {100} series of SnTe and EuS are identified. The atomic spacings of both EuS and SnTe were extracted and found to be respectively 2.98 Å and 3.14 Å, close to the bulk values. Figure 1(c) shows an XTEM image of the EuS layer and the two surrounding SnTe layers. The strain that results from the 5% lattice mismatch between EuS and SnTe is relieved by the formation of several line dislocations that can be at both interfaces (red circles in Fig. 1(c)), but nevertheless, there is excellent epitaxial registration of the lattices. The cube-on-cube alignment of EuS on SnTe (and vice versa) can be confirmed from both the electron diffraction pattern and TEM image. Details of the TEM studies are given in supplement S2.

Samples were then patterned into 1 mm x 0.3 mm Hall bars using standard photolithography. The Hall bar pads were contacted with indium thus allowing electrical contact to both the top and the bottom layers. Transport measurements and SQUID magnetometry were then performed up to 5 T and down to 2 K in a Quantum Design SQUID magnetometer equipped with a modified transport probe. [42]

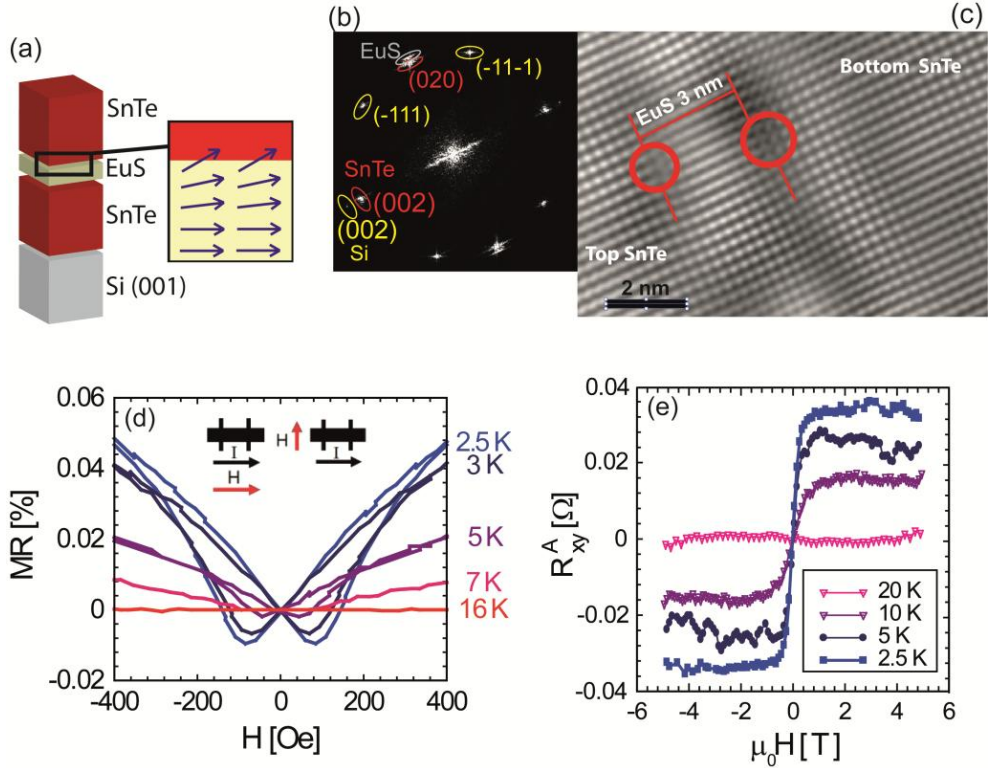

FIG. 1. (Color online) (a) Illustration of the SnTe-EuS-SnTe trilayer device. The magnetic moment distribution at the EuS-SnTe interface is shown on the right. (b) Electron diffraction pattern of the trilayer identifying diffraction spots of EuS, SnTe and Si, respectively in grey, red and yellow. (c) Fourier-filtered cross-sectional TEM image of the trilayer showing line dislocations appearing at EuS/SnTe interface. (see supplement S2 for details). (d) The in-plane MR showing two resistance minima at the coercive field of EuS for temperatures below the Curie temperature of EuS. The behavior was found to be isotropic under a 90° rotation of the magnetic field in the film plane. (e) Anomalous Hall resistance (details in supplement S3) extracted for four temperatures above and below the 16 K Curie point of EuS.

### A. Evidence of induced magnetism from magnetotransport

The magnetotransport measurements were found to exhibit two signatures characteristic of magnetism induced onto the conducting surface-states at the EuS-SnTe interface. *(i) A negative hysteretic in-plane MR is observed at low temperature exhibiting resistance minima at the coercive field of EuS ($H_C$~100 Oe) for two orthogonal in-plane directions of the applied magnetic field*. The minima disappear when the temperature approaches the measured Curie temperature of EuS (16 K) similar to a previous study of $Bi_2Se_3$-EuS bilayers.[3] Further measurements closely correlate the observed transport to the magnetic domain texture of the EuS. These results may be due to the presence of domain-wall-trapped 1D conduction channels at the SnTe-EuS interfaces. *(ii) An anomalous Hall resistivity is observed below the Curie point of EuS despite the fact that EuS is insulating (Fig. 1(e))*. Furthermore, the results of magnetometry as well as the AHE demonstrate that the magnetic moment at the SnTe-EuS interface is canted towards the interface normal direction, as illustrated in Fig. 1(a).

## B. Enhanced conduction at the coercive field

Figure 2(a) shows the in-plane MR=[R(H)-R(0)]/R(0) measured at 2.5 K and 16 K with the magnetic field applied respectively parallel and perpendicular to the current. Pronounced resistance minima appear at the coercive fields for both magnetic field directions.

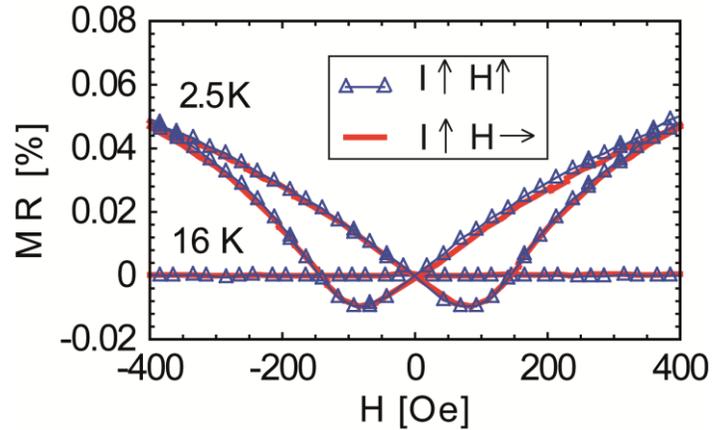

FIG. 2. (Color online) (a) In-plane MR measured with the field oriented either parallel or perpendicular to the current direction at 2.5 K and 16 K.

Since the MR is found to be isotropic when the field is rotated by 90° with respect to the direction of the current, we can rule out the role of spin-scattering that typically results in a strong MR anisotropy in ferromagnetic metals and semiconductors. [43] The fact that a resistance *minimum* is observed instead of a maximum at the coercive field also rules out possible parallel conducting channels that arise in EuS due to charge transfer, as that would result in resistance maxima at the coercive field. [44] Other classical mechanisms such as fringing fields can also be ruled out. (See supplement S4). It is important to note that the MR is clearly hysteretic, developing minima only as the applied field sweeps past H=0 towards ±$H_c$, which is concurrent with the total magnetic moment reaching zero. At this point the FMI layer has the maximum number of magnetic domains and the highest density of domain walls. The enhanced conductivity at the coercive field and isotropic behavior of the MR thus reflect the possible creation of additional 1D conduction channels supported by the domain walls. [1, 3, 7, 31]

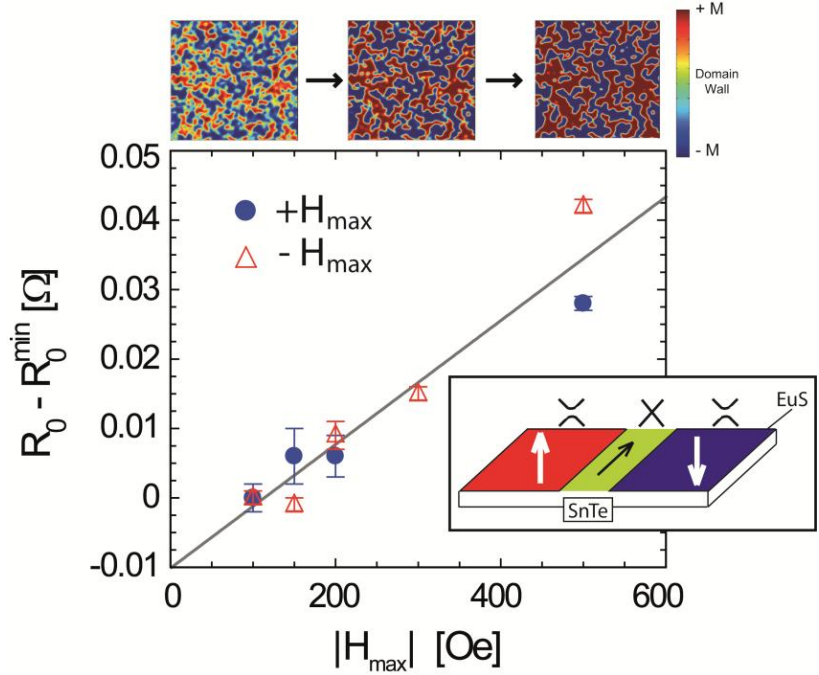

FIG. 3. (Color online) The change in the zero-field resistance $R_0$ plotted as a function of $H_{max}$, the magnetizing field. The line is a guide for the eye. Simulation of the evolution of the magnetic domain texture as a function of the magnetizing field are shown above, where higher $|H_{max}|$ leads to a decrease in domain-wall density. Details of the measurements are described in supplement S5. Inset: Schematic diagram showing a 1D non-trivial conduction channel (green) between two oppositely magnetized magnetic domains of EuS occurring at the EuS-SnTe interface. The white arrows represent perpendicular magnetic moment directions at the interface and the black arrow represents the domain wall conduction direction. (above) The surface state dispersion becomes gapped in the domains on either side of the domain wall due to the perpendicular component of the moment at the interface.

### C. Zero-field resistance as a function of domain-wall density

As the measured magnetotransport may be strongly connected to the formation of magnetic domains, the resistance of the device at *zero applied field* should also exhibit some dependence on the magnetic domain texture. In order to test this, we measured the zero-field resistance $R_0$ after progressively reducing the domain-wall density by magnetizing the EuS layer in a systematically increasing magnetic field $H_{max}$ between 100 and 500 Oe (see details in supplement S5). Figure 3(a) plots the change in the zero-field resistance $\Delta R_0 = R_0 - R_0^{min}$ versus $|H_{max}|$, where $R_0^{min}$ is the resistance when EuS is fully demagnetized. $R_0$ clearly increases for increasing $|H_{max}|$. As the EuS film becomes increasingly magnetized, the domain-wall density decreases as smaller magnetic domains nucleate into larger ones. The nucleation of magnetic domains can be simulated by considering the interaction of classical spins in a hypothetical square array. (See supplement S6 for more details). [36, 45] The evolution of magnetic domain nucleation is illustrated at the top of the plot in Fig. 3. This change in domain-wall density is characteristic of other soft magnetic thin film systems. [46, 47] By analogy to the resistance decrease at the coercive field, a decreasing domain-wall density leads to a decreasing number of domain-wall trapped conduction channels and thus should lead to an increase in the zero-field resistance. This result further correlates the resistivity of the trilayer with the domain-wall density of the FMI.

The fact that the conductivity of the trilayer is enhanced in the presence of the highest density of domain walls may point to the existence of domain-wall-trapped conduction channels at the TCI-FMI interfaces. Such highly-conductive states have been predicted to emerge at the magnetic domain wall, [1, 48] in the presence of a perpendicular magnetic moment required to break time-reversal-symmetry and induce mass terms to the otherwise massless surface Dirac Fermions of a TCI, thus opening up a dispersion gap. [22, 23] At a domain-wall separating two oppositely-oriented magnetic domains, with moments as shown by the vertical arrows in the inset of Fig. 3(c), the perpendicular exchange magnetic field can be zero. This vanishing of magnetic field at the domain wall may be responsible for generating a non-trivial 1D conduction state. [1, 2] At the interface between a TCI and an FMI, the complex magnetic texture of the FMI yields a network of such 1D conduction channels The magnetic domain texture at the EuS-SnTe interface can be radically changed by applying a magnetic field or by magnetizing the EuS layer, thus making a direct impact on this domain-wall supported conduction network. [3, 31] A systematic increase in the magnetizing field for example, was seen to induce an increase in resistance (Fig. 3), a behavior that can be well explained by this hypothesis.

Although we do not have direct causative evidence of the existence of a domain-wall trapped conduction network, our study presents correlative evidence between electron transport and the magnetic domain texture. One downside that prevented a more direct result is the fact that our Fermi level is located about 0.2 eV below the valence band edge of SnTe. The amount of doping in our samples is thus similar to what is typical found in $Bi_2Se_3$ films used in previous works. [3-5]

However, theoretical studies of $Bi_2Se_3$ in proximity to magnetism have shown that such 1D states are likely to disperse far into the conduction and valence bands of the TI. [48] These 1D topological states are expected to be robust and make a small contribution to the conductivity even in moderately doped films with finite Fermi level, much like the contribution of surface states when bulk bands are occupied. [5,31,47] It is also advantageous to note that in the case of the SnTe-EuS interface, if any band bending should occur, it would be expected to deplete carriers in SnTe, since one expects a type-III broken gap band alignment at the interface. [49,50, 51] A 1D conduction network thus provides a reliable explanation for the observed conductivity enhancement at the coercive field (Fig. 2) as well as the behavior of the resistance as a function of the magnetizing field (Fig. 3). The effects are expected to be small in magnitude due parasitic bulk conduction.

### D. Canted interfacial magnetic moment

The resistance changes described above rely on a component of the magnetic moment that is perpendicular to the SnTe-EuS interface. Although the easy axis of the EuS film is primarily in-plane, measurements of the AHE and magnetization reveal that the moment is canted at the interface. Figure 4(a) shows the hysteretic behavior and remanence of the anomalous Hall resistance at 2.5 K. In general, a hard-axis AHE is not expected to show hysteresis. The observation of a remanent Hall effect is evidence of a canted moment at the SnTe interface at zero applied field. Additionally, the AHE plotted in Fig. 4(b) is seen to saturate faster than the out-of-plane magnetic moment. This suggests that the magnetism induced onto the conducting SnTe as measured by the AHE can be more easily aligned by a perpendicular magnetic field than the bulk out-of-plane moment of EuS as measured in SQUID magnetometry. The canted moment is thus expected to be larger at the SnTe-EuS interface.

Finally, we compare the magnetization of the trilayer measured by SQUID magnetometry in both the in-plane and out-of-plane directions. As expected, the easy axis of EuS is dominantly in-plane and the out-of-plane hard axis magnetic moment does not reach saturation until 0.9 T (Fig. 4(c)). The magnetization saturates at about 6$\mu_B$/Eu$^{2+}$ a significantly lower value than what was measured in $Bi_2Se_3$-EuS bilayer; a likely result of interface dislocations and strain. The magnetization in the hard-axis direction exhibits a finite coercivity with a remanence of about 10 % of that measured in the in-plane direction (Fig. 4(d)). We obtain a lower bound of 6° on the canting angle with respect to the horizontal if we assume that the magnetic moment was homogeneously canted throughout the entire thickness of the EuS layer. We can speculate on the origin of this canting. Although it is possible that strain-induced magnetic anisotropy may tilt the moment out of the film plane as in the case of YIG on GGG, [7] the canting may be due to intrinsic spin-orbit interaction effects at the SnTe-EuS interface. [52] In addition, surface anisotropy may contribute a significant component of the anisotropy energy and thus cant the moment out of the plane [46, 53]. Magnetic force microscopy, neutron scattering and theoretical studies will be important tools to understand the origin of such canting.

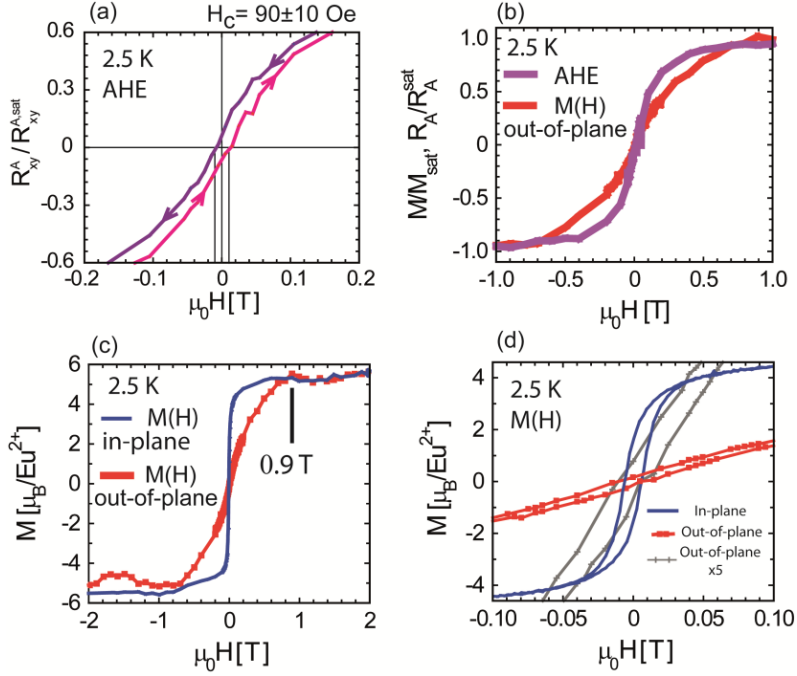

FIG. 4. (Color online) Magnetic field dependence of resistance and magnetization measured at 2.5 K. (a) Hysteresis in the anomalous Hall resistance at low applied field. (b) Comparison of the anomalous Hall resistance to the out-of-plane magnetic moment. The AH resistance saturates at lower magnetic field than the out-of-plane magnetic moment. (c) Magnetic hysteresis loops measured with a magnetic field applied either in the in-plane direction (thin blue curve) or the out-of-plane direction (red curve with points). The magnetic easy axis is in-plane. (d) Magnetic hysteresis loops at low applied fields showing a finite coercive field for both the in-plane (blue curve) and out-of-plane directions (red curve with points and grey curve with crosses), evidence of a canted moment at zero applied field.

## III. CONCLUSIONS

In conclusion, magnetotransport measurements on SnTe-EuS-SnTe trilayer structures reveal the existence of magnetism induced onto the surface of the TCI by the ferromagnetic insulator. The observed hysteretic magnetotransport and zero-field resistance is correlated with the magnetic domain structure, providing stronger evidence for domain-wall-supported 1D conduction. SQUID magnetometry and a remanent AHE provide evidence of a canted magnetic moment at the interface between SnTe and EuS. This perpendicular moment at the SnTe surface would be a necessary condition for a magnetic field breaking time-reversal-symmetry and opening a gap in the surface state dispersion. Since the TCI state provides an inherently tunable platform to study and manipulate topological surface states, these results contribute a crucial ingredient for future pursuits of exotic quantum phenomena arising from proximity-magnetized topological crystalline surface states and facilitate tailoring the properties of such states.

ACKNOWLEDGEMENTS

This research was supported by grants from the National Science Foundation DMR-0907007 and ECCS-1402738 and partly by grants from NSF-DMR-1207469, ONR-N00014-13-1-0301, and the STC Center for Integrated Quantum Materials under NSF-DMR-1231319.